\documentclass[a4paper]{llncs}

\pdfoutput=1 

\usepackage{listings-lean2}
\usepackage{tikz}
\usepackage{multirow}
\usepackage{amsmath,amssymb}
\usepackage{hyperref}
\usepackage[capitalize]{cleveref}
\usepackage[style=base,font=small,labelfont=bf,format=hang,labelsep=period]{caption}
\usepackage{csquotes}
\usepackage{microtype}

\usepackage{xspace}

\newcommand\barrel{BARReL\xspace}

\makeatletter
\newcommand{\rw}[2]{%
	\ifsimplenotes@disable #2%
	\else \fixme[{\normalfont\st{#1} \textcolor{red}{#2}}]%
	\fi%
}
\makeatother

\usetikzlibrary{shapes.misc, positioning, chains, calc}

\newcommand\bminop{\ensuremath{\operatorname{min}}}

\newcommand\bmin[1]{\ensuremath{\bminop(#1)}}

\newcommand\bdom[1]{\ensuremath{\operatorname{dom}(#1)}}
\newcommand\bapp[2]{\ensuremath{#1(#2)}}
\newcommand\binterval[2]{\ensuremath{#1 .. #2}}
\newcommand\bfinone[1]{\ensuremath{\operatorname{FIN}_1({#1})}}

\newcommand\pfun{\mathrel{\ooalign{\hfil$\mapstochar\mkern5mu$\hfil\cr$\to$\cr}}}
\newcommand\pinj{\mathrel{\ooalign{\hfil$\mapstochar\mkern5mu$\hfil\cr$\rightarrowtail$\cr}}}
\newcommand\tinj{\rightarrowtail}
\newcommand\psurj{\mathrel{\ooalign{\hfil$\mapstochar\mkern5mu$\hfil\cr$\twoheadrightarrow$\cr}}}
\newcommand\tsurj{\twoheadrightarrow}
\newcommand\ndres{\mathbin{\rlap{\hbox{$-$}}{\lhd}}}
\newcommand\bovr{%
  \mathbin{\triangleleft
    \mkern-4mu
    \raisebox{0.525ex}{\rule{0.35em}{0.04em}}%
    \mkern-1.3mu}%
}

\newcommand\leanpf{\cdot\mkern-5mu\cdot\mkern-5mu\cdot}

\newcommand\metavar[1]{\ensuremath{?{#1}}}

\colorlet{keywordcolor}{blue!80!white}
\colorlet{symbolcolor}{keywordcolor}
\colorlet{sortcolor}{orange!90!black}
\colorlet{tacticcolor}{keywordcolor}
\colorlet{attributecolor}{black}
\colorlet{commentcolor}{gray}
\colorlet{stringcolor}{green!70!black}

\lstdefinelanguage{b} {
	morekeywords=[1]{MACHINE, REFINEMENT, IMPLEMENTATION, REFINES, SEES, INCLUDES, IMPORTS, EXTENDS, SETS, CONSTANTS, VARIABLES, ASSERTIONS, PROPERTIES, INVARIANT, OPERATIONS, PRE, IF, THEN, END, INITIALISATION, ANY, WHERE},
	morekeywords=[2]{NATURAL, NAT, NATURAL1, NAT1, INTEGER, INT, BOOL, BOOL, POW, POW1, FIN, FIN1, card, min, max},
	morecomment=[l]{//},
	showstringspaces=false,
	mathescape=true,
	keepspaces=true,
	breaklines=true,
	numbers=left,
	numberstyle=\tiny\textcolor{black},
	keywordstyle=[1]\small\itshape\textcolor{keywordcolor},
	keywordstyle=[2]\small\textcolor{sortcolor},
	commentstyle=\textcolor{commentcolor},
	identifierstyle={\ttfamily\textcolor{black}},
	numbersep=6pt,
	columns=[l]fullflexible,
	literate=
		{∈}{{\ensuremath{\in}}}1
		{∉}{{\ensuremath{\notin}}}1
		{∧}{{\ensuremath{\land}}}1
		{∨}{{\ensuremath{\lor}}}1
		{¬}{{\ensuremath{\lnot}}}1
		{⇒}{{\ensuremath{\implies}}}1
		{⇔}{{\ensuremath{\iff}}}1
		{≠}{{\ensuremath{\neq}}}1
		{≤}{{\ensuremath{\leq}}}1
		{≥}{{\ensuremath{\geq}}}1
		{→}{{\ensuremath{\to}}}1
		{↔}{{\ensuremath{\leftrightarrow}}}1
		{∅}{{\ensuremath{\varnothing}}}1
		{∪}{{\ensuremath{\cup}}}1
		{∩}{{\ensuremath{\cap}}}1
		{⊆}{{\ensuremath{\subseteq}}}1
		{⊂}{{\ensuremath{\subset}}}1
		{⇒}{{\ensuremath{\Rightarrow}}}1
		{∀}{{\ensuremath{\forall}}}1
		{∃}{{\ensuremath{\exists}}}1
		{↣}{{\ensuremath{\tinj}}}1,
	extendedchars=true
}

\newcommand\inlineab{\lstinline[language=b, breaklines=true, breakatwhitespace=true, mathescape=false, escapeinside={}, keywordstyle={}, keywordstyle={[2]}, basicstyle=\normalsize\ttfamily]}

\usepackage{etoolbox}
\makeatletter
\patchcmd{\section}{-18\p@}{-14\p@}{}{\errmessage{Section spacing patch failed}}
\patchcmd{\section}{12\p@}{8\p@}{}{\errmessage{Section spacing patch failed}}
\patchcmd{\subsection}{-18\p@}{-12\p@}{}{\errmessage{Subsection spacing patch failed}}
\patchcmd{\subsection}{8\p@}{6\p@}{}{\errmessage{Subsection spacing patch failed}}
\patchcmd{\paragraph}{-12\p@}{-8\p@}{}{\errmessage{Paragraph spacing patch failed}}
\patchcmd{\@listI}{8\p@}{4\p@}{}{\errmessage{List spacing patch failed}}
\let\@listi\@listI
\makeatother
\AtBeginDocument{%
  \setlength{\abovedisplayskip}{7pt plus 2pt minus 2pt}%
  \setlength{\belowdisplayskip}{7pt plus 2pt minus 2pt}%
  \setlength{\belowdisplayshortskip}{4pt plus 2pt minus 2pt}%
}

\title{\barrel: A Modern Backend for Atelier B in Lean}

\author{Ghilain Bergeron\orcidID{0009-0008-2814-3480} \and Vincent Tr\'elat\orcidID{0009-0006-4143-3939}\thanks{ANR project BLaSST (ANR-21-CE25-0010).}}
\institute{Universit\'e de Lorraine, CNRS, Inria, LORIA, Nancy, France}

\authorrunning{G. Bergeron and V. Tr\'elat}

\begin{document}

\maketitle

\thispagestyle{plain}
\begin{abstract}
\barrel is a Lean~4 library bridging Atelier~B, an industrial tool for the B method, and the Lean proof assistant by enabling users to conduct their formal B developments---up to machine refinement and implementation---interactively inside Lean, while retaining standard B syntax.
B partial operators are encoded with explicit well-definedness conditions, using Lean's dependent types to enforce a well-definedness discipline by construction.
That is, proof obligations and proof steps cannot silently rely on ill-typed or ill-defined instantiations.
\barrel also provides basic automation for discharging these conditions.
The implementation is written entirely using Lean metaprogramming and is designed to be modular: extending the supported B fragment typically requires only adding new syntax and encoding clauses.
We compare the resulting workflow with Atelier~B on a B refinement development and examine the checks required during interactive proof through two examples.

	\keywords{Lean~4 \and Atelier~B \and Metaprogramming \and Interactive Theorem Proving}
\end{abstract}

\section{Introduction}

The B method~\cite{Bmethod} and its associated toolsets, notably Atelier~B \cite{AtelierB}, have been used for decades in industrial developments where strong assurance arguments are required.
The B method is a formal development methodology based on stepwise refinement of abstract high-level specifications into executable code, supported by a rich mathematical language rooted in set theory and first-order logic.
B specifications are written as \emph{abstract machines} with sets, constants, variables, properties, invariants, and operations; Atelier~B generates \emph{proof obligations} (POs) whose discharge establishes, for example, invariant preservation and refinement correctness, and provides automated and interactive provers tailored to this logic.
In many certified developments, these tools are trusted as part of the verification chain.

In parallel, interactive theorem provers have seen rapid adoption, driven by their expressive type theories, powerful automation, and growing libraries of reusable mathematics and verification components.
Lean~4~\cite{lean} is based on the calculus of inductive constructions with quotients and has a dependent type system.
It offers a small trusted kernel, a standard library, and a metaprogramming framework that supports its use as both a programming language and a proof assistant.
However, despite B’s long-standing industrial role, support for using modern interactive theorem provers as backends for Atelier~B remains very limited: some previous work targets Isabelle~\cite{IsabelleHOL} but is either no longer maintained~\cite{Schmalz2012RodinIsabelle} or provides specific support~\cite{Wolff2024EB_Isabelle} for Event-B~\cite{AbrialEventB} in Rodin~\cite{AbrialRodin}.
Earlier work also studied a typed functional semantics of B and its translation to PVS~\cite{Bodeveix2002BtoPVS}.

This paper presents \emph{\barrel} (B Automated tRanslation for Reasoning in Lean), a Lean~4 library that bridges Atelier~B and Lean by turning B proof obligations into Lean theorem statements to be proved by automation or interactively.
Given a B machine (or an existing POG file), \barrel invokes Atelier~B to generate POs when needed and translates them into Lean theorems expressed over Mathlib~\cite{mathlib4}'s set-theoretic primitives, preserving the original structures and notations as closely as possible.
\barrel also generates its own \emph{well-definedness} (WD) proof obligations, which capture the definedness side conditions of B partial operators (e.g.,\ minimum, function application, and cardinality).
These are treated as standard Lean statements, so that definedness constraints can be proved once and then reused in subsequent interactive proofs.
The library provides commands for importing B developments and proving their obligations with standard Lean tactic scripts; the resulting theorems are added to the environment under names derived from the original POG tags.
The translation pipeline is organized into clearly separated components: a lightweight embedding of B syntax, a POG file reader, an encoding layer mapping B constructs to Lean expressions, and a discharger that generates theorem declarations and connects them to user proofs. This high-level architecture is illustrated in \cref{fig:barrel-archi}.

\begin{figure}[t]
	\centering
	\begin{tikzpicture}[
			file/.style={chamfered rectangle, draw, chamfered rectangle corners=north west},
			entity/.style={rectangle, draw},
			node distance=1.5cm and 1.5cm,
			every node/.style={inner sep=6pt, align=center, minimum height=1.22cm, outer sep=0},
			arr/.style={->, shorten <=3pt, shorten >=3pt}
		]
		\node[file] (B mch) {.mch};
		\node[file, right=of B mch] (B bxml) {.bxml};
		\node[entity, right=of B bxml] (pos) {Proof\\obligations};
		\node[entity, right=of pos] (thms) {Lean\\theorems};

		\draw[arr] (B mch) -- node[above=-4mm, midway] {\scriptsize Parsing} (B bxml);
		\draw[arr] (B bxml) -- node[above=-6mm, midway] {\scriptsize PO\\[-.6ex]\scriptsize generation} (pos);
		\draw[arr] (pos) -- node[above=-6mm, midway] {\scriptsize PO\\[-.6ex]\scriptsize encoding} (thms);
		\draw[arr, shorten <=7pt, shorten >=7pt] (thms.south) arc[start angle=220, end angle=500, x radius=9.2mm, y radius=9.5mm] node[midway, fill=white, minimum height=0pt, inner sep=2pt] {\scriptsize Proving};
	\end{tikzpicture}
	\caption{High-level picture of \barrel.}
	\label{fig:barrel-archi}
\end{figure}

Our contributions are threefold:
\begin{itemize}
	\item \barrel, an open-source, modular backend for B developments, written in Lean~4;
	\item a translation pipeline preserving B notations and proof obligations, with explicit well-definedness conditions for partial operators;
	\item an analysis of proof checking and modular reuse, combining two Atelier~B examples with a complete B refinement development.
\end{itemize}
Potential extensions include a verified proof obligation generator and an embedded DSL for writing B developments directly in Lean.

The rest of this paper is organized as follows.
After reviewing related work, we recall the necessary background on the B method and present the design and implementation of \barrel, including its treatment of partial operators and basic automation.
We then examine the checks required during interactive proof (\cref{sec:proof-safety}), motivating the comparison of lemma reuse, refinement dependencies, and automation in the leader-election case study (\cref{sec:case-study}).
We conclude with future extensions and the remaining trust assumptions.

\section{Related work}
\label{sec:related}

Several approaches have explored using interactive theorem provers to increase confidence in developments written in the B family of methods by moving proof obligations into a prover with a small, auditable kernel.
An early line of work translates fragments of B into PVS, using type synthesis to recover the typing information needed by the target logic and then relying on PVS for interactive discharge of the resulting obligations~\cite{Bodeveix2002BtoPVS}.
In the Isabelle/HOL ecosystem, mechanizations have focused primarily on Event-B.
The Isabelle plugin for Rodin~\cite{Schmalz2012RodinIsabelle} provides a formalization of the Event-B logic, including treatment of partial functions via definitional extensions and supporting automation.
A more user-facing shallow embedding was proposed recently, viewing Event-B as a DSL hosted in Isabelle/HOL~\cite{Wolff2024EB_Isabelle}.

\barrel is complementary in scope and emphasis.
Rather than formalizing the logic, it targets the existing Atelier~B workflow.
By consuming POG files or invoking the generator, \barrel turns each obligation into a Lean theorem over Mathlib’s set-theoretic primitives~\cite{mathlib4}, while preserving B notations and focusing on usability for B developers.
\barrel provides basic automation to reduce the initial proof effort, with an emphasis on explicit handling of well-definedness conditions.

\section{Overview of \barrel}

This section presents \barrel and the workflow depicted in \cref{fig:barrel-archi}.
We first recall the B method, then explain how \barrel handles well-definedness conditions, and finally detail its inner workings and automation features.

\subsection{Background}

\paragraph*{The B method and proof obligations.}
The B method is a state-based formal development approach centered on \emph{abstract machines} and stepwise \emph{refinement}.
A machine declares abstract sets and constants, a state described by variables constrained by an inductive invariant, an initialization, and operations.
A development proceeds by refining an abstract machine into more concrete refinements down to an implementation, introducing representation data together with a \emph{gluing invariant} relating concrete and abstract states.
From each machine and refinement step, Atelier~B generates \emph{proof obligations} (POs): logical sequents expressing the required invariant-preservation and refinement conditions.
Concretely, POs cover invariant preservation by the initialization and each operation under its precondition, and simulation-style obligations ensuring that concrete operations preserve the abstract behavior through the gluing invariant.
In addition, the PO generator emits \emph{well-definedness} (WD) obligations for the many partial constructs of the language.

POs can be discharged automatically using Atelier~B’s internal provers and by delegating to external backends.
In particular, several approaches translate B obligations to SMT solvers such as Z3~\cite{z3} or cvc5~\cite{cvc5} via dedicated encodings, including the \emph{ppTransSMT}~\cite{DeharbeBSMT} and the more recent \emph{BEer}~\cite{beer} encodings.
Obligations that remain---often those requiring nontrivial quantifier instantiation, refinement reasoning, or additional WD reasoning---are typically discharged interactively using Atelier~B’s GUI-oriented prover, where the user guides rule applications and instantiations.

\paragraph*{The B language in a nutshell.}
B's mathematical language is based on classical set theory and first-order logic with equality.
Specifications are written using the usual set-theoretic constructors such as comprehension, powerset, Cartesian product, and an algebra of relations.
Ordered pairs are first-class values (written as \emph{maplets} $x \mapsto y$), and relations are simply sets of such pairs, i.e.,\ $R \subseteq A \times B$.
Functions are not primitive: a partial function $f$ from $A$ to $B$ (written $f \in A \pfun B$) is a relation $f \subseteq A \times B$ whose domain is contained in $A$ and which is \emph{functional}:
for all $x \in A$ and $y,z \in B$, if $(x \mapsto y) \in f$ and $(x \mapsto z) \in f$ then $y = z$.
Total ($A \longrightarrow B$), injective (partial $A \pinj B$ and total $A \tinj B$), and surjective (partial $A \psurj B$ and total $A \tsurj B$) variants are defined similarly.

Although the underlying logic is set-theoretic and thus untyped, Atelier~B enforces a static well-formedness discipline often described as ``typing'': every identifier and expression must have a deducible ``parent set.''
This guarantees in particular that constructed sets are homogeneous and rules out ill-formed terms.
These deduced parent sets provide exactly the information needed to interpret B expressions in Lean's typed logic in our translation.

\begin{example}
	The expression $1 + \top$ is well-formed in ZFC set theory---and definitionally equal, under the encoding \(\bot \hookrightarrow 0\) and \(\top \hookrightarrow 1\), to the set $\mathbb{B}$ of Booleans, or the natural number $2$---but it is not a valid expression in B.
	Indeed, $1$ has parent set $\mathbb{Z}$ and $\top$ has parent set $\mathbb{B}$, and $+$ expects both its arguments to have parent set $\mathbb{Z}$.

\end{example}

\paragraph*{Foundations of B in Lean.}
We use these typing annotations to ground our encoding in typed higher-order logic (rather than ZFC set theory), using Mathlib's \texttt{Set} and \texttt{SetRel} types\footnote{These are defined, respectively, as $\mathtt{Set}\ \alpha \triangleq \alpha \to \mathtt{Prop}$ and $\mathtt{SetRel}\ \alpha\ \beta \triangleq \mathtt{Set}(\alpha \times \beta)$.} to represent sets and relations:
each inferred B carrier is mapped to a corresponding Lean type---\texttt{INTEGER} to \texttt{Int}, Cartesian products to Lean product types, and powersets to \texttt{Set} over the element type.
This is a convenient representation since Mathlib comes
with many definitions and theorems related to \texttt{Set} for us to use when proving the translated obligations in Lean.
This representation also allows us to express every kind of set used in B specifications since all sets in B are required to be \emph{homogeneous}, i.e.,\ restricted to a uniform shape of elements.
For instance, the set $\{0, \text{``hi''}\}$ must be considered ill-formed by implementations of the B method.

We start by defining standard B operators directly in Lean---reusing their notations from B, in order for the POs to be easily understandable by B users---and typing them appropriately in Lean to keep their use as generic as possible.
Some examples (the set of relations, the set of partial functions, the set of total injective functions, and domain subtraction) are given in \cref{lst:B-sets-lean}.
\begin{lstlisting}[label=lst:B-sets-lean, caption={Encoding of some B operators in Lean.}, float=t]
abbrev rels {α β} (A : Set α) (B : Set β) : Set (SetRel α$$ β) :=
  𝒫$$ (A ×$$ˢ B)
scoped infixl:125 "  ⟷$$  " =>$$ rels

abbrev pfun {α β} (A : Set α) (B : Set β) : Set (SetRel α$$ β) :=
  { f ∈$$ A  ⟷$$  B | ∀$$ ⦃x y z⦄, (x, y) ∈$$ f →$$ (x, z) ∈$$ f →$$ y = z }
scoped infixl:125 " ⇸$$ " =>$$ pfun

abbrev injTFun {α β} (A : Set α) (B : Set β) : Set (SetRel α$$ β) :=
  A ⤔$$ B ∩$$ A  ⟶$$  B
scoped infixl:125 " ↣$$ " =>$$ injTFun

abbrev domSubtr {α β} (E : Set α) (R : SetRel α$$ β) : SetRel α$$ β$$ :=
  { z ∈$$ R | z.1 ∉$$ E }
scoped infixl:160 " ⩤$$ " =>$$ domSubtr
\end{lstlisting}
We then provide proofs, using Mathlib's infrastructure and theorems, of various laws and theorems of B: for any total function $f \in A \longrightarrow B$ its domain $\bdom{f}$ equals $A$; assuming that $a \le b$, the minimum of $\binterval{a}{b}$ is $a$; etc.
Both the definitions and theorems are readily available to end users when importing \barrel in a Lean project, and serve as basic building blocks for proofs of B obligations translated by our tool.

\subsection{Encoding partial B operators}
\label{sec:partial-operators}
A central challenge in this setting is the use of \emph{partial} B operators, meaning that they are only defined when some \emph{well-definedness} (WD) conditions are satisfied.
A typical example is the minimum operator $\bmin{S}$, which returns the smallest element of a set $S$ (of integers or reals) with respect to $\le$, if it exists, and is undefined otherwise.
These partial operators are typically encoded using Hilbert's bounded choice operator $\epsilon\ x \in S.\ P$, which is left unspecified if no element of $S$ satisfies $P$.
The \emph{WD condition} of the corresponding operator is precisely that $P$ is satisfied by at least one element of $S$.

Atelier~B generates separate WD obligations from a component's source clauses.
The principal POs do not carry these obligations as explicit proof arguments; establishing the development's correctness therefore requires relating the separate WD checks to the principal proofs.
This generation does not cover arbitrary terms introduced subsequently by interactive commands.
In the example in \cref{sec:lean-ip-comparison}, an interactive instantiation leads the prover to accept a contradictory assertion.

In \barrel, dependencies between WD goals and principal goals are represented by explicit proof arguments, as in the encoding of Russell~\cite{sozeau:2006} into Rocq.
First, we generate WD conditions to match the exact context in which they are inserted.
Atelier~B's generated WD conditions may differ because they undergo simplification and because our encoding uses different definitions for some predicates and WD conditions.

\begin{remark}
	\barrel defines the WD of minimum as the existence of a least element in the set, and maximum dually as the existence of a greatest element.
	The definitions apply uniformly to partially ordered types.
	For integers, the library proves equivalence with the non-emptiness and finiteness conditions used by Atelier~B: a minimum requires a finite negative part, and a maximum a finite non-negative part.
	For reals, mere non-emptiness and boundedness would not suffice: the interval $(0,1)$ has a lower bound but no least element.
	The integer equivalences are stated as follows:
	\begin{lstlisting}[language=lean]
theorem min.WD_iff_AtelierB_WD {S : Set ℤ} :
	min.WD S ↔$$ S ≠$$ ∅$$ ∧$$ S ∩$$ (INTEGER \ NATURAL) ∈$$ FIN INTEGER
theorem max.WD_iff_AtelierB_WD {S : Set ℤ} :
	max.WD S ↔$$ S ≠$$ ∅$$ ∧$$ S ∩$$ NATURAL ∈$$ FIN INTEGER
\end{lstlisting}
\end{remark}

These equivalences relate the operator conditions; they do not imply a one-to-one correspondence between the generated WD obligations.
Atelier~B may simplify its WD goals and remove hypotheses, whereas \barrel generates conditions in the exact Lean context of each encoded occurrence and shares their proofs when possible.
%
Then, we use Lean's dependent types to encode partial operators using \texttt{Classical.choose},\footnote{Hilbert's $\epsilon$ operator, whose type is \mbox{$\{\alpha : \mathtt{Sort}\ u\} \to \{p : \alpha \to \mathtt{Prop}\} \to (\exists\, x,\, p\ x) \to \alpha$}} which explicitly depends on a proof that the predicate is satisfied by at least one element.\footnote{The encoding represents partial operators as total Lean functions with explicit WD proof parameters.}
This means that one cannot construct a term depending on a partial operator without providing a proof that it is well-defined.
The statement of this proof, which is the WD condition of the operator, is then generated as a side theorem (or a subgoal) which must be discharged for the main theorem to be fully proven, i.e.,\ not depend on the \texttt{sorryAx} axiom---the axiom that is introduced by admitted theorems.
To illustrate this use of \texttt{Classical.choose}, the encoding (and WD conditions) of $\bmin{S}$ and function application $\bapp{F}{x}$ are given in \cref{lst:min-fn-app}.

\begin{lstlisting}[language=lean, mathescape=true, label=lst:min-fn-app, caption={Encoding of the $\bmin{S}$ and $\bapp{F}{x}$ partial operators.}, float=t]
structure min.WD {α} [PartialOrder α] (S : Set α) : Prop where
  isBoundedBelow : ∃$$ x ∈$$ S, ∀$$ y ∈$$ S, x ≤$$ y

noncomputable 
abbrev min {α} [PartialOrder α] (S : Set α) (wd : min.WD S) : α$$ :=
  Classical.choose wd.isBoundedBelow

structure app.WD {α β} (f : SetRel α$$ β) (x : α) : Prop where
  isPartialFunction : f ∈$$ dom f  ⟶$$  ran f
  isInDomain : x ∈$$ dom f -- $\textcolor{commentcolor}{\equiv\ \exists}$ y, (x, y) ∈$$ f

noncomputable 
abbrev app {α β} (f : SetRel α$$ β) (x : α) (wd : app.WD f x) : β$$ :=
  Classical.choose wd.isInDomain
\end{lstlisting}

\subsection{Inner workings of \barrel and automation}
\label{sec:innner-workings}

\paragraph*{User-level workflow.}

\barrel is organized into four distinct passes.
First, the machine, system, refinement, or implementation is parsed into the BXML format, and proof obligations are generated from this format using Atelier~B's internal tools \texttt{bxml} and \texttt{pog}.

Although Atelier~B's \texttt{pog} can generate WD conditions for the given machine, we choose to rely on our own generation instead, since the WD conditions that are generated by \texttt{pog} may not exactly match the context in which they will appear in the Lean terms.
Then, the resulting POs (in XML format) are parsed by \barrel into an abstract syntax tree (AST) and normalized.
Normalization ensures that B predicates like $P \land Q \Rightarrow R$ are treated the same as the logically equivalent predicate $P \Rightarrow Q \Rightarrow R$.
Normalization is not required for the translation, but reduces the number of conjunction-elimination steps needed in the resulting Lean goals.
The normalized B formulas are then transcribed into Lean terms using Lean's facilities for metaprogramming and elaboration.
Encoding partial B operators requires tracking the WD conditions inserted during translation.
We use Lean's metavariables, which retain their local contexts, to represent these conditions.
A WD metavariable is assigned an application of a theorem with the required type: either a suitable theorem already available for reuse (via exact type matching or subsumption), or a fresh obligation extracted from its local context.
Subsumption, here, allows deriving a fact from a \texttt{@[subsume]}-tagged theorem with fewer hypotheses (i.e.,\ a more general statement), e.g.,\ deriving \(P \Rightarrow Q \Rightarrow R\) from \(P \Rightarrow R\), \(Q \Rightarrow R\), or just \(R\).
Proved WD theorems can be reused within an import, but also, once the component is completed, by subsequent imports, including across Lean modules; \cref{sec:proof-workflow} explains the smaller theorem interfaces that make this reuse possible across refinements.
\begin{example}
	We illustrate how \barrel handles WD conditions using the following (already normalized) B predicate, stating that every nonempty finite set of integers---i.e.,\ belonging to $\bfinone{\mathbb{Z}}$---whose elements are nonpositive has a nonpositive minimum: \[
		\forall S \cdot (S \in \bfinone{\mathbb{Z}} \Rightarrow (\forall x \cdot x \in S \Rightarrow x \le 0) \Rightarrow \bmin{S} \le 0)
	\]
	The translation first generates a Lean term while inserting metavariables in place of WD conditions (where $h_1$, $h_2$, and $h_3$ are fresh identifiers):
	\[
		\begin{split}
			 & P :=                                                                                                                                                                    \\
			 & \quad \forall\, (S : \mathtt{Set}\, \mathbb{Z})\, (h_1 : S \in \mathtt{B.Builtins.FIN_1}\ \mathbb{Z})\, (h_2 : \forall\, (x : \mathbb{Z})\, (h_3 : x \in S),\ x \le 0), \\
			 & \qquad\qquad \mathtt{B.Builtins.min}\ S\ (\metavar{m_1}\ S\ h_1\ h_2) \le 0
		\end{split}
	\]
	The new metavariable $\metavar{m_1}$ is generated internally by \barrel and contains $S : \mathtt{Set}\ \mathbb{Z}$, $h_1 : S \in \mathtt{B.Builtins.FIN_1}\ \mathbb{Z}$, and $h_2 : \forall\, (x : \mathbb{Z})\ (h_3 : x \in S),\ x \le 0$ in its local context.
	Although its type is $\mathtt{B.Builtins.min.WD}\ S$, its local context is implicitly universally quantified, hence the full application $\metavar{m_1}\ S\ h_1\ h_2$ in the goal.
	A fresh theorem statement $min\_wd_1$ is then generated---and remains to be proved---from the type of $\metavar{m_1}$ universally quantified over its local context: \[
		\begin{split}
			 & \mathtt{theorem}\ min\_wd_1 :                                                                                                                                          \\
			 & \quad \forall\, (S : \mathtt{Set}\ \mathbb{Z})\, (h_1 : S \in \mathtt{B.Builtins.FIN_1}\ \mathbb{Z})\, (h_2 : \forall\, (x : \mathbb{Z})\, (h_3 : x \in S),\ x \le 0), \\
			 & \qquad\quad \mathtt{B.Builtins.min.WD}\ S
		\end{split}
	\]
	$\metavar{m_1}$ is finally assigned $min\_wd_1$, and substituted in $P$ to obtain the final goal to be proved by the user: \[
		\begin{split}
			\vdash \forall\, & (S : \mathtt{Set}\ \mathbb{Z})\, (h_1 : S \in \mathtt{B.Builtins.FIN_1}\ \mathbb{Z})\, (h_2 : \forall\, (x : \mathbb{Z})\, (h_3 : x \in S),\ x \le 0), \\
			                 & \quad \mathtt{B.Builtins.min}\ S\ (min\_wd_1\ S\ h_1\ h_2) \le 0
		\end{split}
	\]
\end{example}

These steps are performed by the \texttt{import <type> <name> from <folder>} command of \barrel.
The input format is specified by \texttt{<type>}: \texttt{machine}, \texttt{system}, \texttt{refinement}, \texttt{implementation}, or \texttt{pog}.
The argument \texttt{<name>} gives the file name without the extension.
For POG files supplied directly, the conversion to BXML and PO generation are skipped.
Each remaining goal can then be selected with \texttt{next obligation} or, by name, with \texttt{obligation <po>}.
Both commands accept \texttt{from <term>} or \texttt{by <proof>} to prove WD conditions and reconstructed proof obligations.
Finally, the import must be closed with \texttt{qed <name>}, which checks that all obligations have been discharged, and only then adds them to the global context as theorems.
However, \texttt{qed} does not check for dependencies on \texttt{sorryAx}; users must check these themselves, as in other Lean developments.




\paragraph*{Proof automation.}
\label{par:automation}
Before presenting any goals to the user, \barrel runs an automation pass whose primary purpose is to discharge routine side conditions induced by our encoding of B partial operators.
Concretely, the tactic layer repeatedly applies relevant lemmas---backtracking as needed---that relate WD subgoals and available hypotheses.
These helper lemmas provide definitional equations and rewriting and simplification facts for the supported B constructs involving sets, relations, functions, arithmetic, and finiteness. Many generated WD goals then reduce to elementary facts about membership, domain/range inclusion, finiteness, and order bounds and can be closed automatically.
We use Mathlib's facilities to tag helper lemmas with relevant categories---\texttt{wd\_min}, \texttt{wd\_max}, \texttt{wd\_app}, \texttt{wd\_card}, \texttt{wd\_mod}, \texttt{wd\_div}---and its set-theoretic results to discharge the corresponding goals.
These tags also name the tactics run automatically by \barrel's automation layer; users may invoke them manually or extend them with their own lemmas.
For function application, the WD tactic can use a total-function membership lemma tagged \texttt{@[tfun]} together with domain membership; \cref{ex:workflow-update} illustrates this extension on a function update in the case study.

Beyond the core encoding of B primitives, \barrel provides a library of theorems about built-in operators.
Its purpose is twofold:
\begin{enumerate}
	\item provide reusable WD facts that discharge the side goals generated for partial operators;
	\item offer canonical rewrite lemmas that turn B-shaped expressions into ``Mathlib-friendly'' goals.
\end{enumerate}
For instance, in the arithmetic fragment used throughout our running examples, the library includes dedicated WD lemmas for the minimum operation, as shown in \cref{lst:min-wd-lemmas}, each of them being tagged with the relevant theorem category.

\begin{lstlisting}[language=lean, caption={WD lemmas about extrema in \barrel.}, label=lst:min-wd-lemmas, float]
@[wd_min]
theorem min.WD_singleton {α : Type _} [PartialOrder α] {a : α} :
	min.WD {a} :=$$ ⋯
@[wd_min]
theorem min.WD_of_finite {α : Type _} [LinearOrder α] {S A : Set α}
	(h : S ∈ FIN₁ A) : min.WD S :=$$ ⋯
@[wd_min]
theorem min.WD_interval {lo hi : ℤ} (h : lo ≤$$ hi) :
	min.WD (lo..hi) :=$$ ⋯
\end{lstlisting}

The library also provides membership results about finite sets and intervals, together with computational equalities tagged as rewrite rules for automation (\cref{lst:finset-interval-lemmas}). Proofs are replaced by $\leanpf$ for brevity.

\begin{lstlisting}[language=lean, caption={Membership lemmas about finite sets and intervals in \barrel.}, label=lst:finset-interval-lemmas, float]
theorem interval.FIN₁_mem {lo hi : ℤ} (h : lo ≤$$ hi) :
	lo .. hi ∈$$ FIN₁ INTEGER :=$$ ⋯
@[simp]
theorem min.of_singleton {α : Type _} [PartialOrder α] {a : α} :
	min {a} (min.WD_singleton) = a :=$$ ⋯
@[simp]
theorem interval.min_eq {lo hi : Int} (h : lo ≤$$ hi) :
	min (lo .. hi) (min.WD_interval h) = lo :=$$ ⋯
\end{lstlisting}

The remaining obligations---typically invariant preservation and refinement simulation goals---are left as interactive proof tasks.

\section{A Checked and Modular Proof Workflow}
\label{sec:proof-safety}

\barrel should be understood as a \emph{proof-producing backend} for the obligations it emits:
once a proof script succeeds, the corresponding theorem is checked by Lean.
Concretely, the remaining trusted assumptions are:
\begin{itemize}
	\item Lean's kernel, and its standard meta-theory,
	\item the correctness of the external PO generation step performed by Atelier~B (\texttt{bxml}/\texttt{pog}) at \barrel's import phase,
	\item the faithfulness of \barrel's set-theoretic embedding of B constructs and proof obligations to the intended B semantics.
\end{itemize}
What \barrel \emph{does} guarantee by construction is that
\begin{enumerate}
  \item partial B operators cannot be used in generated Lean terms without supplying explicit WD proofs;
  \item any successfully discharged goal yields a kernel-checked Lean theorem---i.e.,\ a Lean (proof-)term having the right type---under that embedding.
\end{enumerate}
A fully independent chain would additionally require a verified PO generator, which we leave as future work.
We examine these two complementary aspects of proof construction described above through examples in Atelier~B.\footnote{Community Edition 24.04.2.}

\subsection{Well-definedness throughout interaction}
\label{sec:lean-ip-comparison}

The WD obligations generated from a B specification concern the expressions occurring in that specification.
Interactive proof can introduce further partial expressions, for instance when choosing an existential witness.
Their definedness must be justified at that point.

Atelier~B's reference manual states that the tactic \texttt{se} does not protect against ill-typing or ill-definedness~\cite[\S4.44]{AtelierBInteractiveProver}.
Consider the machine defined in \texttt{specs/DerivFalse.mch} containing the contradictory assertion \(\exists x \in \mathbb{Z}, x \notin \mathbb{Z}\).
After an initial deduction step (\texttt{dd}), the command \inlineab!se(min({}))! instantiates $x$ with \(\min\varnothing\).
The displayed goal becomes
\[
	\bmin{\varnothing} \in \mathbb{Z}
	\;\land\;
	\bmin{\varnothing} \notin \mathbb{Z}.
\]
A call to \texttt{mp}, the mini-prover, closes this contradictory goal although \(\min\varnothing\) is undefined.

In \barrel, the same instantiation requires a proof of \(\mathtt{min.WD}\ \varnothing\) (\cref{lst:min-fn-app}), that is,
\[
	\exists x \in \varnothing,\ \forall y \in \varnothing,\ x \le y,
\]
which is unprovable.
Admitting this WD condition triggers a Lean warning, and the \texttt{assert\_no\_sorry} check in \texttt{examples/DerivFalse.lean} fails on the resulting theorem.
For operators encoded with WD parameters, definedness is therefore checked throughout proof construction, including steps the source PO generator could not anticipate.

\subsection{Validity of inference rules}
\label{sec:checked-rules}

Inference rule validity is a separate requirement from definedness.
For instance, the rule \texttt{GenPredicateX.8} in Atelier~B's supplied database has the following schema:
\[
	\frac{a \notin \mathcal{P}(X) \cup \{y\}}
	{\exists z\in a,\ z\notin X \land z\ne y}.
\]
The premise places $y$ at the type of $a$, whereas the conclusion compares it with an element of $a$.
The schema therefore has no consistent B typing for arbitrary \(y\).
Yet, replacing \(y\) with the empty set \(\varnothing\) results in a well-typed formula---the two instances of the empty set are of different types---that yields an invalid inference.
%
Indeed, let $a := \{\varnothing\}$ and $X := \mathcal{P}_1(\mathbb{Z})$, the set of nonempty subsets of $\mathbb{Z}$.
Since ${\varnothing \in a}$ but $\varnothing \notin X$, we have $a \notin \mathcal{P}(X)$. Since $a \ne \varnothing$,
the premise $a\notin\mathcal{P}(X)\cup\{\varnothing\}$ holds.
The conclusion requires an element of $\{\varnothing\}$ distinct from $\varnothing$, which is impossible.

The following script derives this contradiction and closes the current goal using only total operators:
\begin{lstlisting}[language=b, basicstyle=\small\ttfamily, numbers=none, mathescape=false, escapeinside={}]
ah(not({{}}: POW(POW1(INTEGER)) \/ {{}})) & pr & dd &
  ar(GenPredicateX.8,Fwd) & pr
\end{lstlisting}

In Lean, we refute this rule instance and prove that replacing \(\mathcal{P}(X) \cup\{y\}\) with $\mathcal{P}(X\cup\{y\})$ in the premise yields a valid rule.
Both proofs are available in \texttt{examples/Falsehood.lean} in the artifacts.

These counterexamples motivate two requirements for a proof workflow: terms introduced during interaction must carry the WD proofs required by their operators, and reusable inference rules must themselves have checked proofs.

\section{Case Study: Comparing the Tools}
\label{sec:proof-workflow}
\label{sec:case-study}

We compare the Atelier~B and \barrel proof workflows on the six-component B development of the IEEE~1394 tree identify protocol by Abrial, Cansell, and M\'ery~\cite{AbrialCansellMery2003}, which refines a leader-election algorithm.
The artifacts contain the B sources in \texttt{specs/leader/} and both sets of proof scripts in \texttt{examples/leader/}, together with the counterexamples from \cref{sec:proof-safety}.
We ran Atelier~B's Force~3 followed by its predicate prover (\texttt{pp}) on the generated obligations, then completed the remaining proofs interactively.

This case study exercises the requirements identified in \cref{sec:proof-safety}: partial expressions carry explicit WD proofs, and the framework can be extended with reusable (kernel-checked) lemmas.
We examine how \barrel supports invariant-preservation proofs after updates, reuse across refinements, and extensions to automation while maintaining these checks.

\subsection{Reusable mathematical lemmas}
\label{sec:workflow-lemmas}

The rule instance examined in \cref{sec:checked-rules} motivates requiring proofs of the general facts used to discharge obligations.
Users can extend \barrel with lemmas proved using Mathlib, through the same mechanism as the built-in library described in \cref{par:automation}.
The function updates in the leader-election development illustrate how one such lemma serves both invariant preservation and definedness proofs.

\begin{example}[Preserving totality after a function update]
\label{ex:workflow-update}
The refinement \texttt{mm3} records the confirmations received at each node in a function $\mathtt{sn} \in \mathtt{ND} \to \mathcal{P}(\mathtt{ND})$.
Its operation \texttt{receive\_cnf} records a confirmation from node $\mathtt{xx}$ at node $\mathtt{yy}$ by updating $\mathtt{sn}(\mathtt{yy})$ to $\mathtt{sn}(\mathtt{yy}) \cup \{\mathtt{xx}\}$.
Preserving the total-function invariant is an instance of the general property
\begin{equation}
  {f \in A \to B} \;\land\; {x \in A} \;\land\; {y \in B}
  \quad\implies\quad
  f \bovr \{x \mapsto y\} \,\in\, A \to B.
  \label{eq:workflow-override}
\end{equation}
Here $f \bovr \{x \mapsto y\}$ denotes the relation obtained by replacing the value of $f$ at $x$ with $y$.
The lemma \texttt{tfun\_\allowbreak override\_\allowbreak singleton} proves \cref{eq:workflow-override} for arbitrary carrier types using Mathlib's sets and relations:
\begin{lstlisting}[language=lean, basicstyle=\small\ttfamily, numbers=none]
@[tfun] theorem tfun_override_singleton {α$$ β$$ : Type*}
  {A : Set α$$} {B : Set β$$} {f : SetRel α$$ β$$} {x : α$$} {y : β$$}
  (hf : f ∈$$ A ⟶$$ B) (hx : x ∈$$ A) (hy : y ∈$$ B) :
    f <+ {(x, y)} ∈$$ A ⟶$$ B :=$$ ⋯
\end{lstlisting}
The proof in \texttt{MM3.lean} applies this lemma with $A := \mathtt{ND}$, $B := \mathcal{P}(\mathtt{ND})$, $x := \mathtt{yy}$, and $y := \mathtt{sn}(\mathtt{yy}) \cup \{\mathtt{xx}\}$.
The operation's guard and the invariant provide the node memberships; the remaining premise is that the updated set is a subset of $\mathtt{ND}$.
\end{example}

The same lemma is applied in \texttt{MM4.lean} to the function \texttt{ba}, which records acknowledgments at each node.
Our Atelier~B scripts establish the corresponding totality claims separately in the contexts of these two obligations.
Atelier~B's related rule \texttt{InFunctionXY.22} establishes preservation of partial-function typing under overriding; totality also requires a domain argument.
Atelier~B supports assertions and user rules, but applying a rule does not require a kernel-checked proof of its correctness.
In Lean, each use of the general lemma supplies its premises and is checked by the kernel.

The totality result also provides the first premise of \texttt{app.WD\_of\_mem\_tfun}.
Together with a proof that the argument belongs to the domain, it establishes the definedness of application to the updated function (\cref{lst:min-fn-app}).
The \texttt{@[tfun]} annotation registers the lemma for total function reasoning, so the WD tactic can perform this composition automatically using the mechanism described in \cref{par:automation}.

\subsection{Refinement with explicit proof dependencies}
\label{sec:refinement-dependencies}

The reuse in \cref{ex:workflow-update} concerns a general mathematical lemma.
Refinement also creates dependencies between proofs of individual components: the WD parameters of encoded operators must be supplied in expressions inherited from an abstract component.
This is the cross-import reuse introduced in \cref{sec:innner-workings}.
When that component has already been proved in Lean, \barrel can supply the required proof arguments by reusing its WD theorems.
Lean checks these applications in the refinement's context, so reuse avoids a new proof obligation without omitting the definedness check.

After \texttt{qed}, a component's WD theorems are available to succeeding imports, including across Lean modules.
While trying to discharge subsequent WD obligations, \barrel tries to reuse these theorems via subsumption, as explained in \cref{sec:innner-workings}.
If no available theorem applies, \barrel generates a new WD obligation.

\begin{example}[Reusing definedness across refinements]
\label{ex:workflow-refinement}
In \texttt{mm1}, an assertion states that when all neighbors of a node $x$ are its children in the current tree $\mathtt{tr}$, the tree assigned to $x$ by $\mathtt{ff}$ is $\mathtt{tr}$.
The refinement \texttt{mm2} inherits this assertion, including the application $\mathtt{ff}(x)$.
Definedness of this application needs only
\[
  \mathtt{ff} \in \mathtt{ND} \to (\mathtt{ND} \pfun \mathtt{ND})
  \quad\land\quad x \in \mathtt{ND}
  \quad\Longrightarrow\quad \mathtt{app.WD}\ \mathtt{ff}\ x.
\]
Atelier~B generates WD obligations for the assertion in \texttt{mm1} and does not generate them again in \texttt{mm2}.
In \barrel, importing \texttt{MM1.lean} into \texttt{MM2.lean} makes the proved WD facts available when encoding the refinement.
All 53 occurrences \texttt{mm2} of the WD above are then justified by applying \texttt{mm1}'s theorem, with no new WD obligation in \texttt{mm2}.

\end{example}

Across the six components, Atelier~B generates 60 WD obligations from the B specification.
\barrel instead extracts WD conditions while translating individual expressions and shares their proofs between occurrences.
It generates 31 WD goals, all proved automatically, and avoids 7 further goals by reusing three previously proved theorems across components.
These goal counts exclude repeated theorem applications, such as the 53 in \texttt{mm2}; every encoded partial application still carries a WD proof.

\subsection{Automation, interaction, and trust}

Atelier~B's automation, developed specifically for B over more than twenty years, leaves fewer obligations for interactive proof in this development.
The examples in \cref{sec:proof-safety} motivate considering both automation coverage and how individual proof steps are checked.
In \barrel, Lean's kernel checks the resulting proof terms and the proofs of every lemma they use.

Interactive proof can also help extend trustworthy automation in \barrel.
In \cref{ex:workflow-update}, the proved totality lemma supplies a reusable step for the WD tactic; \cref{ex:workflow-refinement} shows how previously proved WD theorems can be reused across imports.
With these generic lemmas and extended premise search, \barrel automatically discharges all locally generated WD goals in the leader-election development.
Adding a tactic or lemma does not add a trusted inference rule: its use must produce a proof term accepted by Lean.

The proof environment also affects the effort needed when automation fails.
Atelier~B displays goals, hypotheses, and proof histories; the message ``The Predicate Prover failed to prove the current goal'' does not distinguish missing premises from unsuccessful search.
Lean's editor integration helps users investigate such failures: a side panel shows the proof state alongside the script, while hover information and navigation to declarations give access to the relevant lemmas and definitions.

Using \barrel requires B users to learn Lean and the relevant Mathlib abstractions, which may be a barrier to adoption. In return, Lean checks the proofs produced by automation and the lemmas they use, including WD proofs reused across components.

\section{Discussion and future work}

The core tooling of \barrel consists of about 2{,}800 lines of code (LoC), excluding blank and comment lines and including the embedding of the B syntax, the parser, the encoder, and the discharger/\allowbreak metaprogramming layer.

The artifacts also include the library of auxiliary equational and simplification lemmas: the B ``builtins'' library (\texttt{Barrel/Builtins/*}) accounts for about 1{,}500 LoC and contains supporting facts about sets, relations, functions, arithmetic, and well-definedness, together with the automation tactics.

\begin{table}[t]
	\caption{Implementation metrics of the artifacts containing \barrel.}
	\label{tab:implementation-metrics}
	\centering
	\begin{tabular}{l@{\kern5em}r}
		\textbf{Component}    & \textbf{LoC}               \\
		\hline
		Core tooling          & \({\sim}\)2{,}800          \\
		Lemmas and automation & \({\sim}\)1{,}500          \\
		Examples and tests    & \({\sim}\)2{,}700          \\
		\hline
		\textbf{Total}        & \({\sim}\)\textbf{7{,}000} \\
	\end{tabular}
\end{table}

The artifacts contain all examples from the paper and additional tests of \barrel's workflow.
These metrics are summarized in \cref{tab:implementation-metrics}.

\paragraph*{Future work.}

We identify two main directions for future work:
\begin{itemize}
	\item Enriching automation beyond WD goals for recurring proof-obligation patterns.
	\item Developing a fully verified proof obligation generator on top of \barrel.
\end{itemize}

\subsection{Automation beyond WD obligations}\label{sec:future-auto}

The current tactic layer of \barrel is intentionally narrow: it targets WD side conditions and leaves the remaining obligations---typically invariant preservation and refinement simulation---to interactive proof.
In practice, these non-WD obligations also exhibit recurring structures---such as routine set/relational algebra, standard refinement simulation patterns, and repeated use of the same invariants---suggesting that they also admit dedicated, domain-specific automation.
Since \barrel operates entirely inside Lean, this automation is \emph{proof-producing}: tactics synthesize proof terms that are checked by the kernel.
The relevant design goal is to increase proof-search power while keeping proofs readable and predictable.

A concrete plan for extending \barrel's automation layer includes:
\begin{itemize}
	\item enriching the simplification and rewriting library for B-style set and relational algebra (domain/range laws, restriction/subtraction, relational image, and composition) so that obligations normalize to ``Mathlib-shaped'' goals;
	\item adding small, focused procedures for functional-relational reasoning (e.g.,\ determinism, domain/range inclusion, extensionality, and application lemmas);
	\item providing lemmas for common PO families and supporting delegation to standard Mathlib tactics.
\end{itemize}
This would allow \barrel to discharge a larger fraction of proof obligations automatically, while keeping the overall architecture modular and maintainable.
A modular design could also expose a generic interface for set and relational reasoning, allowing users to supply their own implementations or strategies.

\subsection{Toward a fully verified backend}
\label{sec:verified-backend}

Although \barrel is a proof-producing backend for the obligations it emits, it currently relies on Atelier~B's \texttt{bxml} and \texttt{pog} binaries for parsing B machines and generating proof obligations.
A natural next step is therefore to implement, \emph{in Lean}, a PO generator for the supported fragment that produces Lean propositions directly over \barrel's embedding---with correct handling of WD conditions---and to establish soundness guarantees stating that discharging these obligations entails invariant preservation and refinement correctness.
First, this would eliminate the main external component of the trusted computing base, thereby yielding a more self-contained verification chain.
Second, this would support an embedded DSL for writing B developments directly in Lean.
Working with Lean terms would allow WD conditions to be included directly in B developments and, where possible, inferred automatically.

A further direction is to prove the soundness of the translation from B obligations to their Lean counterparts.
While the current translation is intentionally syntax-directed and close to a one-to-one mapping into our embedding, making this correspondence explicit would further reduce the amount of trusted code in the backend, although this would require a significant formalization effort---including giving formal semantics to a fragment of Lean.

\section{Conclusion}
We presented \barrel, a Lean~4 backend for Atelier~B that turns industrial B artifacts into ordinary Lean goals over a set-theoretic encoding of the B language.
Our central design choice is to make B's ubiquitous partiality explicit: partial operators are represented as total Lean functions parameterized by proofs of well-definedness which are reified as standard additional goals.
The requirement applies both to expressions translated from the specification and to terms introduced during interactive proof.

The first Atelier~B example closes a contradictory assertion using \(\min\varnothing\); the second does so through a supplied rule whose counterinstance uses only total operators.
In \barrel, the former requires an impossible definedness proof; Lean refutes the latter rule instance using Mathlib.
The leader-election development illustrates how the same proof discipline supports reuse: mathematical lemmas isolate arguments from generated contexts, and inherited WD conditions are justified by explicit applications of earlier components' theorems.
Automation can use these checked facts without becoming an additional trusted inference mechanism.

Lean's kernel checks the resulting arguments, but the external PO generator and the faithfulness of the translation remain trusted.
Future work includes extending the supported B fragment and lemma library, improving the interface to generated hypotheses, and establishing an end-to-end guarantee through a verified translation and proof obligation generator.

\paragraph*{Artifacts.}
A snapshot of \barrel is available at \url{https://anonymous.4open.science/r/BARReL}.
The archive contains the complete Lean~4 implementation of \barrel (B surface embedding, POG reader, encoder, discharger macros, and tactic layer), together with the sample B machines and Lean scripts used in the evaluation.


\bibliography{ref}
\end{document}